\documentstyle[12pt]{article}
\title{Discrete version of the Chazy class III equation}
\author{Simon Labrunie and Robert Conte\\
        Service de physique de l'\'etat condens\'e\\
        CEA Saclay\\
        91191 Gif-sur-Yvette cedex (France)\\
        e-mail: {\tt labrun@spec.saclay.cea.fr}}

\newcommand{\s}{\sigma}

\newcommand{\g}{\gamma}
\renewcommand{\l}{\lambda}
\newcommand{\eps}{\varepsilon}

\newcommand{\ol}{\overline}
\newcommand{\oll}[1]{\overline{\overline#1}}
\newcommand{\ul}{\underline}
\newcommand{\ull}[1]{\underline{\underline#1}}

\newcommand{\eqn}{equation}
\newcommand{\Eqn}{Equation}
\newcommand{\dif}{differential}
\newcommand{\deqn}{\dif\ \eqn}
\newcommand{\fdeqn}{finite-difference \eqn}
\newcommand{\sq}{following}

\newcommand{\st}{such that}

\begin{document}
\maketitle

\begin{abstract}
We study the discretisation of the Chazy class III equation by two means: a discrete Painlev\'e test, and the preservation of a two-parameter solution to the continuous equation. We get that way a best discretisation scheme.
\end{abstract}

The Chazy class III \eqn~\cite{chy}
\begin{equation}
C\equiv u''' - 2\,u\,u'' + 3\,u'^2 = 0
\label{cha}
\end{equation}
is \st\ the only singularity of its general solution is a movable noncritical natural boundary, a circle whose center and radius depend on the three initial conditions of the Cauchy problem. Thus, \Eqn~(\ref{cha}) has the Painlev\'e property.

\medbreak

Our aim is the obtention of a ``most faithful'' finite-difference representation of this \eqn. To do so, we shall explain first our method of getting discretising  ans\"atze for \deqn s. Then, we will demand that the discrete \eqn\ possesses the ``discrete Painlev\'e property'' defined in the sense of R.~Conte and M.~Musette~\cite{cmu}. Last, we will examine the preservation of a two-parameter solution to \Eqn~(\ref{cha}).

\section{Faithful discretisation of an ODE}

Given a (continuous) $N$-th order \deqn
\begin{equation}
\forall x,\quad{\cal E}(x,u(x),\dots,u^{(N)}(x)) = 0
\label{gen-ed}
\end{equation}
of degree $m$ in $u^{(N)}$, a discretisation scheme may be called a {\it (faithful) discrete version\/} of~(\ref{gen-ed}) if it satisfies the four conditions:
\begin{enumerate} 
\item\label{dem1} It is an $N$-th order \fdeqn, i.e. an iteration relation between $N+1$ values of the unknown function $u$ taken at points in an arithmetic sequence: 
\begin{equation}
\forall x,\,\forall h,\quad E(x,h,u(x + k_0\,h),\dots,u(x+ (k_0 + N)\,h) ) = 0
\label{gen-fd}
\end{equation}
The parameter $h$ is dubbed the {\it step\/} of the \fdeqn; the constant $k_0$ is an origin whose utility will become clear later.
\item\label{dem2} {\it(If \Eqn~(\ref{gen-ed}) has the Painlev\'e property)}\\
It is of degree $m$ in the first and last term, $u(x + k_0\,h)$ and $u(x+ (k_0 + N)\,h)$, whose presence is dictated by the highest-order derivative.
\item\label{dem3} It is invariant by $(k_0,h)\mapsto(-k_0,-h)$. The reason for this condition is that the step $h$ in~\Eqn~(\ref{gen-fd}) can be arbitrarily chosen in a neighbourhood of the origin, and hence be changed for its opposite.
\item\label{dem4} Naturally, its continuous limit ($h\to0$) is \Eqn~(\ref{gen-ed}).
\end{enumerate}
We shall refer to these conditions as the {\it naive discretisation rules\/}. The reason for this terminology is that they simply ensure a formal, easily noticeable, similarity between a \deqn\ and its discretisation scheme. They do {\it not\/} imply the preservation of the distinctive features of the continuous \eqn, such as the Painlev\'e property, linearisability, analytic expression of the general solution, etc.

\medbreak

In the case of \Eqn~(\ref{cha}), we have a third-order \deqn\ which is of first degree in the third-order derivative. Hence, a faithful discrete version of  \Eqn~(\ref{cha}) is a relation between $\ull u = u(x-3\,h/2)$, $\ul u = u(x-h/2)$, $\ol u = u(x+h/2)$, $\oll u = u(x+3\,h/2)$, of first degree in $\oll u$ and $\ull u$, and invariant by $h\mapsto-h$, $\ol u\mapsto\ul u$, $\oll u\mapsto\ull u$.

\smallbreak

There is only one discrete version of the term $u'''$ satisfying these conditions:
$$h^{-3}\,(\oll u - 3\,\ol u + 3\,\ul u - \ull u)$$
but there are {\it a priori\/} three linearly independent discrete equivalents of $u\,u''$; hence this term will be discretised as
$$\matrix{ &\l_1\,h^{-2}\,\left( \ol u\,( \oll u - 2\,\ol u + \ul u ) + ( \ol u - 2\,\ul u + \ull u )\,\ul u \right)/2\cr
	   +&\l_2\,h^{-2}\,\left( \ul u\,( \oll u - 2\,\ol u + \ul u ) + ( \ol u - 2\,\ul u + \ull u )\,\ol u \right)/2\cr
	   +&\l_3\,h^{-2}\,\left( \ull u\,( \oll u - 2\,\ol u + \ul u ) + ( \ol u - 2\,\ul u + \ull u )\,\oll u \right)/2\cr}$$
with $\l_1 + \l_2 + \l_3 =1$. Similarly, the term $u'^2$ possesses three valid discrete equivalents, and will be discretised as
$$\mu_1\,h^{-2}\,(\ol u - \ul u)^2 + \mu_2\,h^{-2}\,(\oll u - \ol u)\,(\ul u - \ull u) + \mu_3\,h^{-2}\,(\oll u - \ul u)\,(\ol u - \ull u)/4$$
with $\mu_1 + \mu_2 + \mu_3 =1$. We thus obtain an expression $E$ whose continuous limit is $C$, the left-hand side of \Eqn~(\ref{cha}). We notice that $E$ depends solely on the two parameters $$\mu'_1 = \mu_1 - {1\over3}\,(2\,\l_1 + 3\,\l_2) + {2\over9},\quad\mu'_2 = \mu_2 - {1\over3}\,(\l_2 - 2\l_1)$$
namely the naive discretisation of~(\ref{cha}) is
\begin{equation}
\matrix{E &\equiv& \ h^{-3}\,\left(\oll u - 3\,\ol u + 3\,\ul u - \ull u\right)\hfill\cr
&+&  {1\over12}\,h^{-2}\,\left( 16\,(\oll u\,\ul u + \ol u\,\ull u) - 3\,\ol u\,\ul u - 27\,\oll u\,\ull u - \oll u\,\ol u - \ul u\,\ull u \right)\cr
&+& {3\over4}\,\mu'_1\,h^{-2}\,\left( 4\,(\oll u\,\ul u + \ol u\,\ull u ) - 3\,\ol u\,\ul u - 3\,\oll u\,\ull u - \oll u\,\ol u - \ull u\,\ul u \right)\cr
&+& {3\over4}\,\mu'_2\,h^{-2}\,\left( \oll u\,\ull u + 4\,(\ol u^2 + \ul u^2) - 7\,\ol u\,\ul u - \oll u\,\ol u - \ull u\,\ul u \right)\hfil=0\cr
}
\label{naiv:d-cha}
\end{equation}

\section{Discrete Painlev\'e test}

Consider an $N$-th order \fdeqn\ like \Eqn~(\ref{gen-fd}), depending on a step $h$. We say this \eqn\ has the {\it (discrete) Painlev\'e property\/} iff its general solution $x\mapsto u(x;h\dots)$ is free from movable critical points in the $x$ plane, provided $h$ belongs to a suitable neighbourhood of the origin.

\medbreak

Suppose that the \fdeqn\ under consideration possesses a continuous ($h\to0$) limit like \Eqn~(\ref{gen-ed}). Then we can apply to it a discrete Painlev\'e test, called {\it method of perturbation of the continuous limit\/}. It was originally set up by R.~Conte and M.~Musette in~\cite{cmu}, and is an analogue of the perturbative Painlev\'e test for continuous \eqn s~\cite{cfp}.

\smallbreak

This test consists in making a perturbative expansion of the general solution $u$ of~(\ref{gen-fd}) in function of an {\it a priori\/} extraneous parameter $\eps$. This induces a similar expansion of $E$:
$$u = \sum_{n=0}^{+\infty} \eps^n\,u^{(n)},\quad E = \sum_{n=0}^{+\infty} \eps^n\,E^{(n)}$$
which has the \sq\ property: all \eqn s $E^{(n)}=0,\ n\ge1$ are the same linear \eqn\ with different right-hand sides, namely
$$E^{(n)}\equiv\left\langle dE^{(0)}, u^{(n)} \right\rangle + R^{(n)}(u^{(0)},\dots u^{(n-1)}) = 0$$
where $\left\langle dE^{(0)}, u^{(n)}\right\rangle$ denotes the differential (or G\^ateaux derivative) of $E^{(0)}$, taken at $u=u^{(0)}$, acting on the test function $u^{(n)}$.

\smallbreak

The choice $\eps=h$ has the extra property that $E^{(0)}={\cal E}$, the continuous limit of $E$. Then, a necessary condition for \Eqn~(\ref{gen-fd}) to possess the Painlev\'e property is that the general solution $u^{(n)}$ of every \eqn\ $E^{(n)}=0$ be free from movable critical singularities.

Practically, look for all possible Laurent series representations
$$u = \sum_{n=0}^{+\infty} \eps^n\,\sum_{j=\rho\,n}^{+\infty} u^{(n)}_j\,\chi^{j+p},\quad \chi=x-x_0$$
where $\rho$ is the least Fuchs index of the linearised zeroth-order \eqn\ $dE^{(0)} = 0$. In this expansion a free parameter, $u^{(n)}_j$, enters at each order $n$ of perturbation every time $j$ is a Fuchs index of $dE^{(0)}$; and the Painlev\'e property implies that all the corresponding coefficients $E^{(n)}_j$ of $\chi^{j+p}$ in $E^{(n)}$ are zero.

\medbreak

We have applied this test to \Eqn~(\ref{naiv:d-cha}). At order~$0$, we get \Eqn~(\ref{cha})
$$E^{(0)}\equiv {u^{(0)}}''' - 2\,u^{(0)}\,{u^{(0)}}'' + 3\,\left({u^{(0)}}'\right)^2 = 0$$
which admits the solution $u=-6\,\chi^{-1}$. Then the Fuchs indices of
$$dE^{(0)}\equiv\partial^3 - 2\,u^{(0)}\,\partial^2 - 2\,{u^{(0)}}''\,{\bf 1} + 6\,{u^{(0)}}'\,\partial $$
are $-3$, $-2$, $-1$. At perturbation order one, we have $E^{(1)}\equiv\left\langle dE^{(0)}, u^{(1)} \right\rangle =0$; whose solution is chosen as $u^{(1)}= \left( u^{(1)}_{-3}\,\chi^{-3} + u^{(1)}_{-2}\,\chi^{-2} \right)\,\chi^{-1}$.

\smallbreak

At perturbation order two, we get the condition
$$E^{(2)}_{-2}\equiv\mu'_2=0$$
Then, if this condition is satisfied, the general solution of \Eqn~(\ref{naiv:d-cha}) is free from movable critical singularities up to perturbation order~$16$ at least, and most probably up to infinity. Thus, there are great chances that \Eqn~(\ref{naiv:d-cha}) has the Painlev\'e property when $\mu'_2=0$.

\medbreak

This condition $\mu'_2=0$ is also the only one given by the singularity confinement criterion of B.~Grammaticos {\it et al.\/} (see~\cite{gramram} for this test).

\section{Two-parameter solutions}

Singularity analysis has given us a one-parameter family of acceptable discrete versions of \Eqn~({\ref{cha}), namely \Eqn s~({\ref{naiv:d-cha}) with $\mu'_2=0$. But the examination of the more specific properties of \Eqn~({\ref{cha}) leads us to restrict our choice, because not all \Eqn s~({\ref{naiv:d-cha}) preserve these features even if $\mu'_2=0$.

\medbreak

For instance, \Eqn~({\ref{cha}) admits a two-parameter particular solution
\begin{equation}
u(x) = -6\,{x-c_1\over (x-c_2)^2}
\label{solpar}
\end{equation}
with $c_1$, $c_2$ arbitrary in the complex plane. Demanding the preservation of the solution~(\ref{solpar}) whatever $c_1$ and $c_2$ yields that the two conditions $\mu'_1=0$ {\it and\/} $\mu'_2=0$ should be satisfied. Hence, the most faithful discretisation scheme is
\begin{equation}
\matrix{\g &\equiv& \ h^{-3}\,\left(\oll u - 3\,\ol u + 3\,\ul u - \ull u\right)\hfill\cr
&+&  {1\over12}\,h^{-2}\,\left( 16\,(\oll u\,\ul u + \ol u\,\ull u) - 3\,\ol u\,\ul u - 27\,\oll u\,\ull u - \oll u\,\ol u - \ul u\,\ull u \right) =0\cr}
\label{d-cha}
\end{equation}

\medbreak

Eliminating $c_1$ and $c_2$ between $u(x)$ calculated by~(\ref{solpar}) and its first two derivatives yields the least-degree second-order \deqn\ satisfied by~(\ref{solpar}):
\begin{equation}
S\equiv 9\,u''^2 + 2\,( u^2 - 9\,u' )\,u\,u'' + 3\,(8\,u' - u^2)\,u'^2 = 0
\label{scha}
\end{equation}
The link between \Eqn s~(\ref{cha}) and~(\ref{scha}) is
\begin{equation}
S' - 2\,u\,S = 2\,(u^3 - 9\,u\,u' + 9\,u'')\,C
\label{link}
\end{equation}
and (\ref{solpar}) is not solution to $u^3 - 9\,u\,u' + 9\,u''=0$.

\medbreak

Let us examine how we can discretise faithfully \Eqn s~(\ref{scha}) and~(\ref{link}). By the naive discretisation rules, a discrete version of \Eqn~(\ref{scha}) should be a relation between $u=u(x)$, $\ol u=u(x+h)$ and $\ul u = u(x-h)$ of second degree in $\ol u$ and $\ul u$ and invariant by $h\mapsto-h$, $\ol u\mapsto\ul u$.

Eliminating $c_1$ and $c_2$ between $u$, $\ol u$ and $\ul u$ calculated by \Eqn~(\ref{solpar}), we get the \sq\ infinite-order discretisation scheme 
\begin{equation}
\matrix{
&9\,h^{-4}\,(\ol u - 2\,u + \ul u)^2 + 3\,h^{-3}\,(\ol u - \ul u)\, (2\,u^2 + u\,(\ol u + \ul u) - 4\,\ol u\,\ul u)\cr
& + h^{-2}\,\left( {1\over4}\,u^2\,(\ol u^2 + \ul u^2) - 2\,u\,\ol u\,\ul u\,(\ol u + \ul u) + 4\,\ol u^2\,\ul u^2 - {1\over2}\,u^2\,\ol u\,\ul u\right) = 0\cr}
\label{d-scha}
\end{equation}
whose general solution is automatically~(\ref{solpar}). We check that \Eqn~(\ref{d-scha}) satisfies the naive discretisation rules.

\medbreak

As for the discretisation of \Eqn~(\ref{link}), we must pay attention to the fact that \Eqn s~(\ref{d-cha}) and~(\ref{d-scha}) do not involve the same values of $u$, despite the misleading use of similar notations. To transpose \Eqn~(\ref{d-scha}) into the world of third-order \fdeqn s, we can think of two  possibilities:
\begin{itemize}
\item $\ol\s$ defined as the left-hand side of \Eqn~(\ref{d-scha}) shifted to the right by half a step, i.e.~formally $\ol u\mapsto\oll u$, $u\mapsto\ol u$, $\ul u\mapsto\ul u$.
\item $\ul\s$ defined as the left-hand side of \Eqn~(\ref{d-scha}) shifted to the left by half a step, i.e.~formally $\ol u\mapsto\ol u$, $u\mapsto\ul u$, $\ul u\mapsto\ull u$.
\end{itemize}
Then the four-point discretisation of the left-hand side $S' - 2\,u\,S$ of \Eqn~(\ref{link}) must obey the naive scheme
$$\omega=(\ol\s - \ul\s)/h - \l_1 (\ol u\,\ol\s + \ul u\,\ul\s) - \l_2 (\ul u\,\ol\s + \ol u\,\ul\s) - \l_3 (\oll u\,\ol\s + \ull u\,\ul\s) - \l_4 (\ull u\,\ol\s + \oll u\,\ul\s)$$
with $\sum_1^4 \l_i=1$. A necessary condition for $\omega$ having the left-hand side $\g$ of \Eqn~(\ref{scha}) as a factor is that the resultant of $\g$ and $\omega$ --- both being seen as polynomials in $h^{-1}$ --- be zero. This happens if and only if
$$(\l_1,\l_2,\l_3,\l_4) = (-{4\over3},{1\over12},0,{9\over4})$$
In that case, we check that $\omega = f\,\g$, with the factor
$$\matrix{f &= {1\over4}\,\left( \oll u\,\ol u^2 + 16\,(\oll u\,\ul u^2 + \ol u\,\ull u^2) - 8\,\ol u\,\ul u\,(\oll u + \ull u) - 5\,\ol u\,\ul u\,(\ol u + \ul u) + \ul u^2\,\ull u \right)\cr
 &+ 3\,h^{-1}\,\left( \oll u\,\ol u - \ol u^2 - 4\,(\oll u\,\ul u - \ol u\,\ull u) + \ul u^2 - \ul u\,\ull u \right)\hfill\cr
 &+ 9\,h^{-2}\,\left( \oll u - \ol u - \ul u + \ull u \right)\hfill\cr}$$
having as its continuous limit $2\,(u^3 - 9\,u\,u' + 9\,u'')$, i.e.~the proportionality factor between $S' - 2\,u\,S$ and $C$ in \Eqn~(\ref{link}).

\section{Conclusion.}

The example treated in this article has shown the efficiency of the method of perturbation of the continuous limit. It has recovered the same condition as that given by the singularity confinement criterion. While the confinement test seems essentially discrete, this perturbative method admits as its continuous limit the continuous Painlev\'e test.

\smallbreak

The integration of the Chazy \eqn\ in terms of solutions to the (linear) hypergeometric \eqn\ was performed by Chazy~\cite{chy} and Bureau~\cite{bur72}. For the discrete analogue~(\ref{d-cha}), this remains an open problem. Indeed, the integration process of the continuous \eqn\ involves an exchange of the dependent and independent variables, a feature which seems hard to transpose into the discrete world.

Then the perservation of the two-parameter solution~(\ref{solpar}) appears as a minimal demand. It leaves only one possibility, which thus may be called the ``most faithful'' \fdeqn\ representing the Chazy \eqn.

\end{document}